\documentclass[11pt,twoside]{article}
\usepackage{asp2014}
\usepackage{hyperref}
\usepackage{footmisc}

\aspSuppressVolSlug
\resetcounters

\begin{document}
\title{MeerKATHI - an end-to-end data reduction pipeline for MeerKAT and other radio telescopes}
\author{MeerKATHI collaboration}
%
\author{Gyula I. G. J\'ozsa$^{1,2,3}$ (delegate), Sarah V. White$^{1,2}$, Kshitij Thorat$^{2,1,4}$, Oleg M. Smirnov$^{2,1}$, Paolo Serra$^{5}$, Mpati Ramatsoku$^{2,5}$, Athanaseus J. T. Ramaila$^{1}$, Simon J. Perkins$^{1}$, D\'{a}niel Cs. Moln\'{a}r$^{5}$, Sphesihle Makhathini$^{2}$, Filippo M. Maccagni$^{5}$, Dane Kleiner$^{5}$, Peter Kamphuis$^{6}$, Benjamin V. Hugo$^{1,2}$, W. J. G. de Blok$^{7,8,9}$, and Lexy A. L. Andati$^{2}$}
\affil{$^{1}$South African Radio Astronomy Observatory, Cape Town, Western Cape, South Africa}
\affil{$^{2}$Department of Physics and Electronics, Rhodes University, Makhanda,  Eastern Cape, South Africa}
\affil{$^{3}$Argelander-Institut f\"ur Astronomie, Auf dem H\"ugel 71, Bonn, Germany}
\affil{$^{4}$Department of Physics, University of Pretoria, Pretoria, Gauteng, South Africa}
\affil{$^{5}$INAF - Osservatorio Astronomico di Cagliari, Selargius, Cagliari, Italy}
\affil{$^{6}$Ruhr-Universit\"at Bochum, Faculty of Physics and Astronomy, Astronomical Institute, Bochum, Germany}
\affil{$^{7}$Netherlands Institute for Radio Astronomy (ASTRON), Dwingeloo, The Netherlands}
\affil{$^{8}$Dept.\ of Astronomy, Univ.\ of Cape Town, Cape Town,  Western Cape, South Africa}
\affil{$^{9}$Kapteyn Astronomical Institute, University of Groningen, Groningen, The Netherlands}


\paperauthor{Gyula I. G.~J\'ozsa}      {jozsa@ska.ac.za}{0000-0003-0608-6258}{South African Radio Astronomy Observatory}               {Cape Town}{Western Cape}  {7925}   {South Africa}\\
\paperauthor{Sarah V. White}           {sarahwhite.astro@gmail.com}        {0000-0002-2340-8303}{Department of Physics and Electronics, Rhodes University}{Makhanda} {Eastern Cape}  {6140}   {South Africa}\\
\paperauthor{Kshitij Thorat} {0000-0002-4760-080X}          {thorat.k@gmail.com}                {}{Department of Physics and Electronics, Rhodes University}{Makhanda} {Eastern Cape}  {6140}   {South Africa}
\paperauthor{Oleg M. Smirnov}          {osmirnov@gmail.com}                {}{Department of Physics and Electronics, Rhodes University}{Makhanda} {Eastern Cape}  {6140}   {South Africa}\\
\paperauthor{Paolo Serra}              {paolo.serra@inaf.it}               {}{INAF - Osservatorio Astronomico di Cagliari}             {Selargius}{Cagliari}      {I-09047}{Italy}\\
\paperauthor{Mpati Ramatsoku}          {mpati.ramatsoku@inaf.it}           {}{Department of Physics and Electronics, Rhodes University}{Makhanda} {Eastern Cape}  {6140}   {South Africa}\\
\paperauthor{Athanaseus J. T. Ramaila} {aramaila@ska.ac.za}                {}{South African Radio Astronomy Observatory}               {Cape Town}{Western Cape}  {7925}   {South Africa}\\
\paperauthor{Simon J. Perkins}            {sperkins@ska.ac.za}{0000-0002-3623-0938}{South African Radio Astronomy Observatory}               {Cape Town}{Western Cape}  {7925}   {South Africa}\\
\paperauthor{Filippo M. Maccagni}      {filippo.maccagni@inaf.it}          {}{INAF - Osservatorio Astronomico di Cagliari}             {Selargius}{Cagliari}      {I-09047}{Italy}\\
\paperauthor{Sphesihle Makhathini}     {sphemakh@gmail.com}                {}{Department of Physics and Electronics, Rhodes University}{Makhanda} {Eastern Cape}  {6140}   {South Africa}\\
\paperauthor{D\'{a}niel Cs. Moln\'{a}r}{daniel.molnar@inaf.it}             {}{INAF - Osservatorio Astronomico di Cagliari}             {Selargius}{Cagliari}      {I-09047}{Italy}\\
\paperauthor{Peter Kamphuis}           {pkamp@astro.ruhr-uni-bochum.de}    {}{Ruhr-Universit\"at Bochum, Faculty of Physics and Astronomy, Astronomical Institute}{Bochum} {Nordrhein-Westfahlen}{44780}{Germany}\\
\paperauthor{Dane Kleiner}             {dane.kleiner@inaf.it}              {https://orcid.org/0000-0002-7573-555X
}{INAF - Osservatorio Astronomico di Cagliari}             {Selargius}{Cagliari}      {I-09047}{Italy}\\
\paperauthor{Benjamin V. Hugo}         {bhugo@ska.ac.za}                   {}{South African Radio Astronomy Observatory}               {Cape Town}{Western Cape}  {7925}   {South Africa}\\
\paperauthor{W. J. G. de Blok}         {blok@astron.nl}                    {}{Netherlands Institute for Radio Astronomy (ASTRON)}      {Dwingeloo}{Drenthe}       {7990AA} {The Netherlands}\\
\paperauthor{Lexy A. L. Andati}        {andatilexy@gmail.com}              {}{Department of Physics and Electronics, Rhodes University}{Makhanda} {Eastern Cape}  {6140}   {South Africa}\\
\paperauthor{Gyula I. G.~J\'ozsa}      {jozsa@ska.ac.za}{0000-0003-0608-6258}{South African Radio Astronomy Observatory}               {Cape Town}{Western Cape}  {7925}   {South Africa}\\
\paperauthor{Sarah V. White}           {sarahwhite.astro@gmail.com}        {0000-0002-2340-8303}{Department of Physics and Electronics, Rhodes University}{Makhanda} {Eastern Cape}  {6140}   {South Africa}\\
\paperauthor{Kshitij Thorat} {0000-0002-4760-080X}          {thorat.k@gmail.com}                {}{Department of Physics and Electronics, Rhodes University}{Makhanda} {Eastern Cape}  {6140}   {South Africa}
\paperauthor{Oleg M. Smirnov}          {osmirnov@gmail.com}                {}{Department of Physics and Electronics, Rhodes University}{Makhanda} {Eastern Cape}  {6140}   {South Africa}\\
\paperauthor{Paolo Serra}              {paolo.serra@inaf.it}               {}{INAF - Osservatorio Astronomico di Cagliari}             {Selargius}{Cagliari}      {I-09047}{Italy}\\
\paperauthor{Mpati Ramatsoku}          {mpati.ramatsoku@inaf.it}           {}{Department of Physics and Electronics, Rhodes University}{Makhanda} {Eastern Cape}  {6140}   {South Africa}\\
\paperauthor{Athanaseus J. T. Ramaila} {aramaila@ska.ac.za}                {}{South African Radio Astronomy Observatory}               {Cape Town}{Western Cape}  {7925}   {South Africa}\\
\paperauthor{Simon J. Perkins}            {sperkins@ska.ac.za}{0000-0002-3623-0938}{South African Radio Astronomy Observatory}               {Cape Town}{Western Cape}  {7925}   {South Africa}\\
\paperauthor{Filippo M. Maccagni}      {filippo.maccagni@inaf.it}          {}{INAF - Osservatorio Astronomico di Cagliari}             {Selargius}{Cagliari}      {I-09047}{Italy}\\
\paperauthor{Sphesihle Makhathini}     {sphemakh@gmail.com}                {}{Department of Physics and Electronics, Rhodes University}{Makhanda} {Eastern Cape}  {6140}   {South Africa}\\
\paperauthor{D\'{a}niel Cs. Moln\'{a}r}{daniel.molnar@inaf.it}             {}{INAF - Osservatorio Astronomico di Cagliari}             {Selargius}{Cagliari}      {I-09047}{Italy}\\
\paperauthor{Peter Kamphuis}           {pkamp@astro.ruhr-uni-bochum.de}    {}{Ruhr-Universit\"at Bochum, Faculty of Physics and Astronomy, Astronomical Institute}{Bochum} {Nordrhein-Westfahlen}{44780}{Germany}\\
\paperauthor{Dane Kleiner}             {dane.kleiner@inaf.it}              {https://orcid.org/0000-0002-7573-555X
}{INAF - Osservatorio Astronomico di Cagliari}             {Selargius}{Cagliari}      {I-09047}{Italy}\\
\paperauthor{Benjamin V. Hugo}         {bhugo@ska.ac.za}                   {}{South African Radio Astronomy Observatory}               {Cape Town}{Western Cape}  {7925}   {South Africa}\\
\paperauthor{W. J. G. de Blok}         {blok@astron.nl}                    {}{Netherlands Institute for Radio Astronomy (ASTRON)}      {Dwingeloo}{Drenthe}       {7990AA} {The Netherlands}\\
\paperauthor{Lexy A. L. Andati}        {andatilexy@gmail.com}              {}{Department of Physics and Electronics, Rhodes University}{Makhanda} {Eastern Cape}  {6140}   {South Africa}\\




  
\begin{abstract}
{\sc MeerKATHI} is the current development name for a radio-interferometric data reduction pipeline, assembled by an international collaboration. We create a publicly available end-to-end continuum- and line imaging pipeline for MeerKAT and other radio telescopes. We implement advanced techniques that are suitable for producing high-dynamic-range continuum images and spectroscopic data cubes. Using containerization, our pipeline is platform-independent. Furthermore, we are applying a standardized approach for using a number of different of advanced software suites, partly developed within our group. We aim to use distributed computing approaches throughout our pipeline to enable the user to reduce larger data sets like those provided by radio telescopes such as MeerKAT. The pipeline also delivers a set of imaging quality metrics that give the user the opportunity to efficiently assess the data quality.
\end{abstract}
\section{Introduction}
The data rate and the size of radio-interferometric data sets has increased to such a level that it is impossible to apply very interactive data reduction strategies, in which the user intervenes frequently in the data reduction process. Instead, automated pipelines are becoming a requirement, in which all data reduction steps (i.e. editing, calibration, imaging, and data quality assessment) are integrated in a single pipeline, with automated data analysis and archiving of data products being a potential extension. 
Such a pipeline should ideally be able to process the data in data reduction loops, which will be controlled by an automated data assessment, reducing the necessity for intervention by the user to a minimum. 

The usage of a data reduction pipeline is often restricted to a specific instrument or a certain software package. While this approach eases the implementation, it is in many cases not the best solution. Optimally, a radio-interferometric data reduction pipeline would make use of the most recent software, based on the most modern techniques, and work regardless of the radiointerferometer providing the data or the origin of the software. 

This is the goal of our international {\sc MeerKATHI} collaboration in building the {\sc MeerKATHI} pipeline. While {\sc MeerKATHI} originates from an effort to create an imaging pipeline suited for the reduction of data generated by the MeerKAT radio telescope, for the purpose of \ion{H}{i} imaging, the focus shifted from this more-restricted case to its application as a generic pipeline. This empowers the user to automatically reduce data from instruments similar to MeerKAT (like the Very Large Array, VLA, or the Giant Metrewave Radio Telescope, GMRT). 
In the following, we describe how the pipeline is implemented, its current scope, first results using the pipeline, and future plans.
\section{Architecture}
{\sc MeerKATHI} is a (large) Python script making use of {\sc Stimela} (the IsiZulu word for a train), a platform-independent radio interferometry scripting framework based on Python and a choice of Linux containerization technologies. It enables the user to use a suite of astronomical software to reduce radio astronomical imaging data. Both software and containerization packages are listed in Table~\ref{tab:repositories}. In addition to containerized (and therefore) platform-independent versions of the software, it provides a standard syntax, allowing the user to access all implemented software in a similar manner. Stimela supports the containerization platforms 
{\sc Podman}, 
{\sc Docker},
{\sc Singularity}, and
{\sc uDocker}.

{\sc Stimela} provides access to standard software as {\sc CASA} \citep[generic data reduction and analysis]{mcmullin_casa_2007}, {\sc PyBDSM} (\citealt{mohan_pybdsf:_2015}, see also \citealt{van_weeren_first_2012}, source finding), {\sc AOFlagger} \citep[radio interference detection and elimination]{offringa_aoflagger:_2010}, {\sc MeqTrees} \citep[data simulation and calibration]{noordam_meqtrees_2010}, {\sc WSClean} \citep[imaging and deconvolution]{offringa_wsclean:_2014}, and {\sc SoFiA} \citep[source finding]{serra_sofia:_2015}. In addition, it wraps high-end software developed within our group, using the same standard syntax:
\begin{itemize}
\setlength\itemsep{-0.5em}
    \item {\sc CUBIcal} (Kenyon et al. 2018, prediction and calibration)
    \item {\sc AImFAsT} (Diagnostics/flow control)
    \item {\sc RFInder} (RFI visualisation) 
    \item {\sc RadioPADRE} (Results examiner)
    \item {\sc RAGaVi} (Data visualisation)
    \item {\sc Tricolour\footnote{\label{note1}Copyright SDP / RARG, 2018-2019. South African Radio Astronomy Observatory (SARAO)}} (Parallel flagger)
    \item {\sc Sunblocker} (Solar RFI mitigation)
    \item {\sc Crystalball} (Parallel predict)
    \item {\sc SHARPener} (Spectral analysis)
    \item {\sc Codex Africanus} (parallel radio software API)
\end{itemize}
\begin{table}[ht]
    \centering
    \begin{small}
    \begin{tabular}{l l c}
   \multicolumn{2}{c}{Software repositories}\\
           \hline
       \hline
	Software & Repository & Reference \\
       \hline
            {\sc MeerKATHI}         &   \url{https://meerkathi.readthedocs.io}                  &  \\
			{\sc AImFAsT} 	    	&	\url{https://github.com/Athanaseus/aimfast/}            &  \\
			{\sc Codex Africanus}   &	\url{https://github.com/ska-sa/codex-africanus/}        &  \\
			{\sc Crystalball}       &	\url{https://github.com/paoloserra/crystalball/}        &  \\
			{\sc CUBIcal} 	    	&	\url{https://github.com/ratt-ru/CubiCal/}               & (1) \\
			{\sc MeqTrees} 	        &	\url{http://meqtrees.net/}                              & (2) \\
			{\sc RadioPADRE}    	&	\url{https://github.com/ratt-ru/radiopadre/}            &  \\
			{\sc RAGaVi} 	    	&	\url{https://github.com/ratt-ru/ragavi/}                &  \\
			{\sc RFInder} 	    	&	\url{https://github.com/Fil8/RFInder/}                  &  \\
			{\sc SHARPener}     	&	\url{https://github.com/Fil8/SHARPener/}                &  \\
			{\sc SoFiA} 	    	&	\url{https://github.com/SoFiA-Admin/SoFiA/}             & (3) \\
			{\sc Sunblocker}    	&	\url{https://github.com/gigjozsa/sunblocker/}           &  \\
			{\sc Tricolour}\footref{note1}     	&   \url{https://github.com/ska-sa/tricolour/}              &  \\
       \hline
			{\sc AOFlagger}	    	&   \url{https://sourceforge.net/p/aoflagger/wiki/Home/}    & (4) \\
			{\sc CASA} 		    	&	\url{https://casa.nrao.edu/}                            & (5) \\
			{\sc PyBDSM} 	    	&	\url{https://github.com/lofar-astron/PyBDSF/}           & (6) \\
			{\sc WSClean} 	    	&	\url{https://github.com/lofar-astron/PyBDSF/}           & (7) \\
       \hline
			{\sc Docker} 	    	&	\url{https://www.docker.com}                            &  \\
            {\sc Podman}            &   \url{https://podman.io}                                 &  \\
			{\sc Singularity}   	&	\url{https://sylabs.io/singularity/}                    &  \\
			{\sc uDocker} 	    	&	\url{https://github.com/indigo-dc/udocker/}             &  \\
       \hline
    \end{tabular}
    \end{small}
    \caption{Repositories of software used in {\sc MeerKATHI}. Top: software with contributions by {\sc MeerKATHI} members. Middle: other radio-astronomical software. Bottom: containerization software. References: (1) \citet{kenyon_cubical_2018}, (2) \citet{noordam_meqtrees_2010}, (3) \citet{serra_sofia:_2015}, (4) \citet{offringa_aoflagger:_2010}, (5) \citet{mcmullin_casa_2007}, (6) \citet{mohan_pybdsf:_2015}, (7) \citet{offringa_wsclean:_2014}.}
    \label{tab:repositories}
\end{table}
\section{Work flow}
\articlefigure{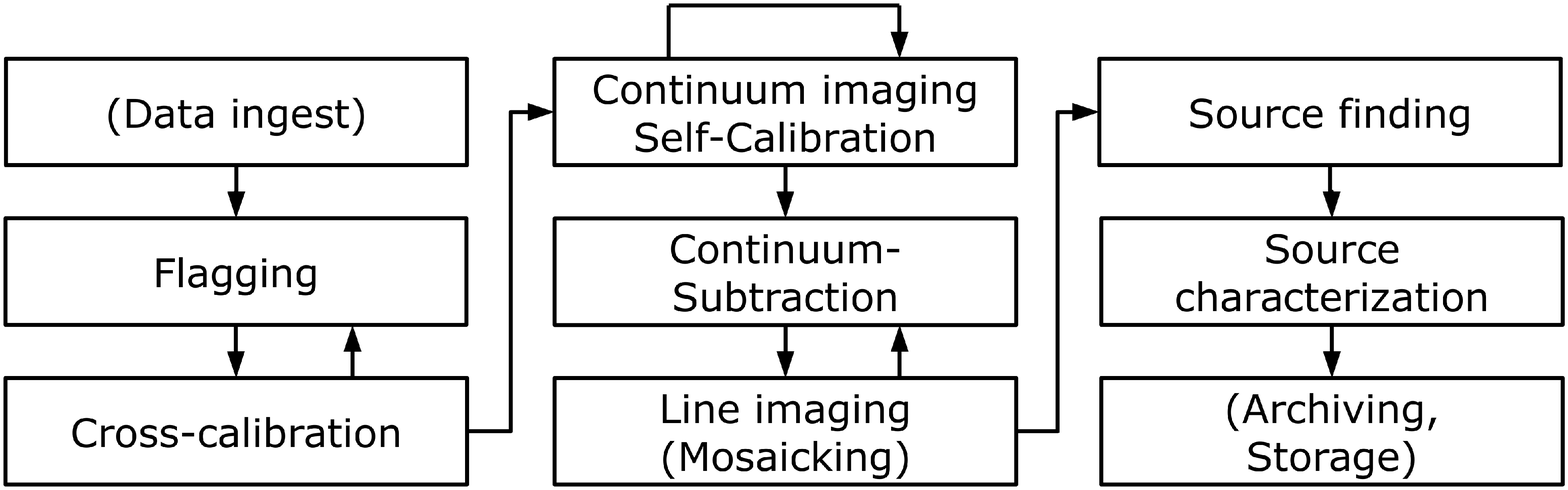}{P10-40_01_workflow}{Exemple {\sc MeerKATHI} workflow.}
{\sc MeerKATHI} has a flexible layout and gets configured via a configuration file. Its simplest function is to flag RFI and to perform cross-calibration (using calibration sources), followed by an iterative series of continuum imaging and self-calibration steps (reducing image artifacts), line imaging, and source finding. Automated data ingest (depending on the telescope), several stages of source characterization, and archiving are planned, but not yet implemented. Figure~\ref{P10-40_01_workflow} shows an example work flow.
A suite of fast, interactive, remotely accessible diagnostic tools has been developed to assess data quality and to identify potential problems. 
\section{Status and Outlook}
After a development phase of less than two years, {\sc MeerKATHI} is able to generate science-ready output from multiple instruments \citep[MeerKAT, VLA, GMRT -- see][]{serra_neutral_2019, ramatsoku_gasp_2019, michalowski_nature_2019, jozsa_2019, maccagni_2019}. We interpret this as an indication that {\sc MeerKATHI} can become a widely used method to reduce radio-interferometric data. While distributed computing is not completely implemented, MeerKATHI is able to process relatively large data sets of several TB. A public release of the software is currently pending verification and documentation, but we give interested users access on a shared-risk policy.

Our attempt to create a radio-interferometric data reduction pipeline, which can process data from multiple instruments using a large suite of available software, has so far been successful. With {\sc MeerKATHI} we provide a working pipeline, which is now available upon request, but will be made public in the near future. Further development will concentrate on improved techniques, distributed computing, automated ingest, analysis, and archiving.
~
\bibliography{P10-40}
\small\noindent {\bf Acknowledgements:} The MeerKATHI team acknowledges support from the Starting Grant 679629 "FORNAX" of the European Research Council (ERC), MAECI Grant ZA18GR02 from the Italian Ministry of Foreign Affairs and International Cooperation, Grant Number 113121 of the South African National Research Foundation as part of the ISARP Joint Research Scheme, and the BMBF project 05A17PC2.

\end{document}